# TREES AS FILTERS OF RADIOACTIVE FALLOUT FROM THE CHERNOBYL ACCIDENT

James D. Brownridge and Noel K. Yeh

Department of Physics, Applied Physics and Astronomy
State University of New York at Binghamton Binghamton,
NY 13902, U.S.A.

This paper is a copy of an unpublished study of the filtering effect of red maple trees (*acer rubrum*) on fission product fallout near Binghamton, NY, USA following the 1986 Chernobyl accident. The conclusions of this work may offer some insight into what is happening in the forests exposed to fallout from the Fukushima Daiichi Nuclear Plant accident. This posting is in memory of Noel K. Yeh.

In an investigation of the radioactive fallout from the Chernobyl accident, it is found that a red maple tree (*acer rubrum*) served as an effective filter for Cs-137. Based on the absorption of C-137 by bark of other trees, the conclusion is drawn that trees in forested watersheds serve as an effective moderator in the contamination of surface water supplies by long-lived radionuclides.

In the course of studying radioactivity among trees, stemflow, and rain, we have accumulated several years of data, some of which have been published. The accident at Chernobyl provided a unique opportunity for this investigation to study the effects of forest trees on the radioactivity of rain water reaching the forest floor, and ultimately, flowing into streams and surface reservoirs.
Our experimental station is located 5 km southeast of the City of Binghamton, NY, and the rain collector is placed in an open area about 100 meters from an 80-year-old 26-m tall red maple tree, from which stemflow was collected at 1 m above ground level. These samples were analyzed by gamma ray spectrometry using an ultra-low-background detector.[1]

The first traces of radioactive fallout in rain water, from a brief shower on May 6, 1986, showed the presence of I-131 and the cosmogenic Be-7. Subsequent data from 15 pairs of stemflow and rain samples, collected simultaneously between May 16, 1986 and July 12, 1986, were analyzed in terms of the "filtering efficiency" **K** of the red maple in removing radioactive nuclides from the rain, where **K** is defined as:

$$K = \frac{\text{Radioactive Concentration in Rain Water} - \text{Radioactive Concentration in Stemflow}}{\text{Radioactive Concentration in Rain Water} + \text{Radioactive Concentration in Stemflow}}$$

The results are presented in Figure 1.

The filtering of naturally occurring Be-7 was relatively constant, unaffected by its concentration in the rain or the type of rainfall. The concentration of Be-7 in the rain varied from approximately $5.4 \times 10^{-2}$ to $5.6 \times 10^{-1}$ Bq/L, and the large positive values of **K** indicate that most of the Be-7 remained on the red maple tree. On the other hand, Figure 1 shows a drastic variation in **K** for Cs-137 between Samples 5 and 14, whose rain water had similar concentration of the isotope. This fluctuation reflects the complex chemical and physical interactions between Cs-137 containing host compounds and the tree. Once the Cs-137 concentration in the rain fell below some critical level, approximately $3.3 \times 10^{-3}$ Bq/L, the filtering action seemed to cease, and the tree became a source of Cs-137 for the stemflow.

Figure 1 also shows that the filtering efficiency fluctuated considerably for I-131 and Ru-103, with most samples exhibiting a negative value of **K**. For much of the time, the red maple tree served as a source of these isotopes after their early deposition. In addition to the radioactive half-lives (8.04 days for I-131 and 39.6 days for Ru-103), we surmise, partly on the basis of on-going studies involving the eastern hop hornbean (*ostrya virginiana*), that another significant parameter is the period of retention of each of these radionuclides by the red maple tree[2]. We conjecture that these periods are shorter than their respective radioactive half-lives for these two nuclides. Note that in Sample 1 (of May 16, 1986); the tree did filter both isotopes because of their initial absence prior to the Chernobyl fallout.

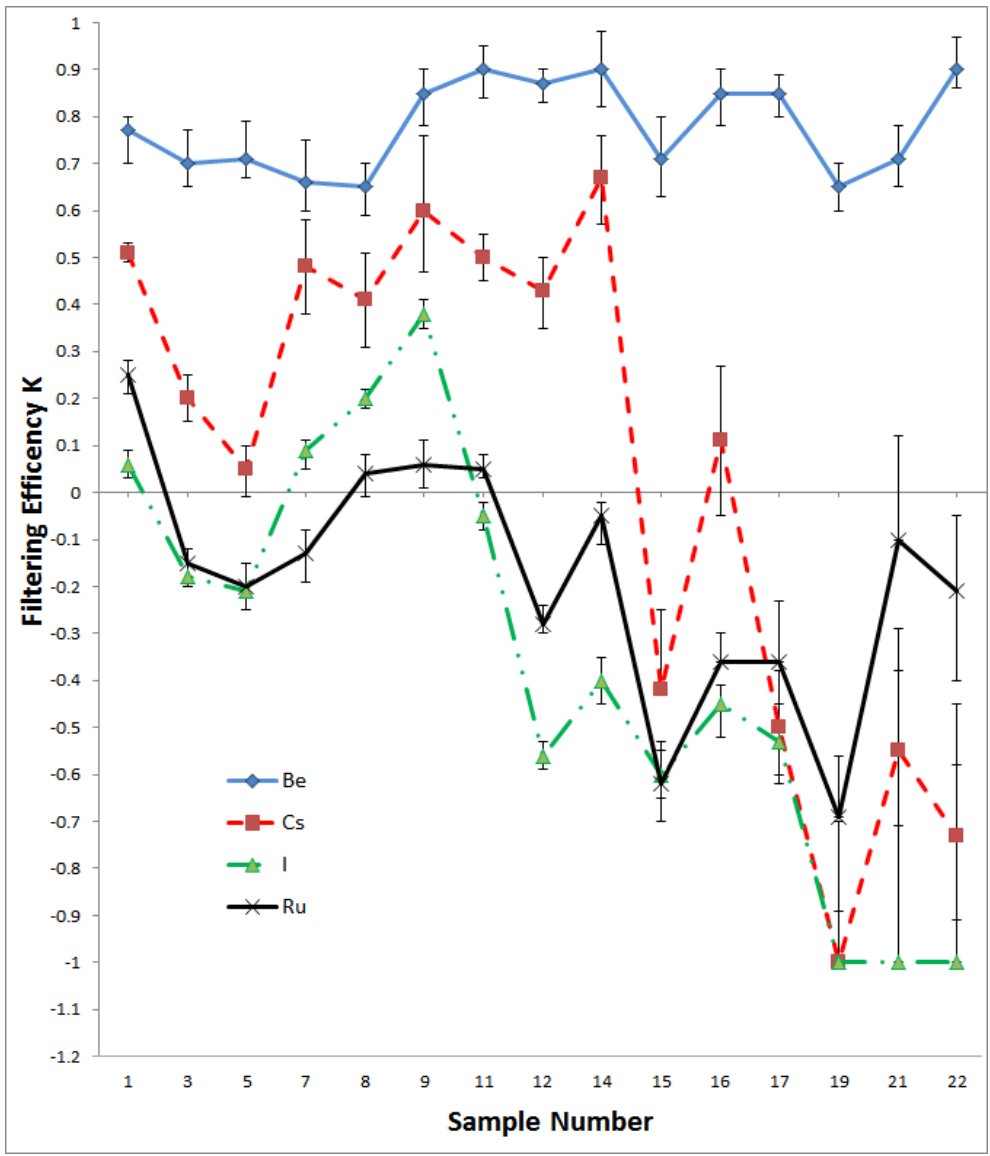

**Figure 1.** Distribution of the filtering efficiency **K** for four radionuclides among samples of stemflow collected from a red maple tree. Only those samples, numbered chronologically, for which there was simultaneous collection of rain samples are plotted. The types of rainfall ranged from drizzles for samples 15-17; intermittent showers for Samples 7, 8, 11, 12, 19-22; to steady rains for Samples 3, 5, 9; and heavy thundershowers for Samples 1 and 14. See text for the definition of **K**. The indicated errors, some of which are half drawn to avoid clutter, include statistical uncertainties only. The lines joining the data points are to facilitate the identification of the different nuclides.

It is apparent from these data that the red maple tree served as both a sink and a source for radionuclides. The stemflow data for this tree, collected randomly during similar, seasons of 1984 and 1985, showed no significant change in the concentration of Cs-137, signifying that the tree has a period of retention for Cs-137 which is much longer than the duration of the present study. The typical stemflow prior to Chernobyl had $(1.0 \pm 0.1) \times 10^{-3}$ Bq/L, with the rain samples showing no detectable levels of Cs-137. There was therefore a continuous transfer of Cs-137 from the red maple bark to the stemflow. In contrast, the Cs-137 level in the stemflow of Sample 1 was $(8.78 \pm 0.62) \times 10^{-3}$ Bq/L.

Averaged over the duration of this investigation, the filtering efficiency **K** of the red maple tree was -0.28 for I-131, -0.18 for Ru-103, +0.13 for Cs-137, and +0.75 for Be-7. As other species of trees also show evidence of retaining Cs-137 in their bark, it is reasonable to conclude that forest trees in general provide an effective filtering of Cs-137 transported from the atmosphere to the forest floor through rain water. This filtering action, especially for forested watersheds that feed surface reservoirs, may be significant in moderating the contamination of water supplies by atmospheric radioactive fallout.